# Fluorescence Reference Target Quantitative Analysis Library

Eammon A. Littler[1], Emmanuel A. Mannoh[1], Ethan P. M. LaRochelle[1*]

QUEL Imaging, White River Junction, VT 05001 USA

*Corresponding author: ethan@QUELimaging.com

## Abstract

Standardized performance evaluation of fluorescence imaging systems remains a critical unmet need in the field of fluorescence-guided surgery (FGS). While the American Association of Physicists in Medicine (AAPM) TG311 report and recent FDA draft guidance provide recommended metrics for system characterization, practical tools for extracting these metrics remain limited, inconsistent, and often inaccessible. We present **QUEL-QAL**, an open-source Python library designed to streamline and standardize the quantitative analysis of fluorescence images using solid reference targets. The library provides a modular, reproducible workflow that includes region of interest (ROI) detection, statistical analysis, and visualization capabilities. QUEL-QAL supports key metrics such as response linearity, limit of detection, depth sensitivity, and spatial resolution, in alignment with regulatory and academic guidance. Built on widely adopted Python packages, the library is designed to be extensible, enabling users to adapt it to novel target designs and analysis protocols. By promoting transparency, reproducibility, and regulatory alignment, QUEL-QAL offers a foundational tool to support standardized benchmarking and accelerate the development and evaluation of fluorescence imaging systems.

## Introduction

Fluorescence-guided surgery was first cleared by the FDA in 2005, and since then, its applications have expanded significantly. New indications for this technology continue to emerge, and recent FDA approvals for fluorescent contrast agents targeting cancers have further advanced the field. Despite this progress, standardization has lagged behind. Imaging device developers seeking FDA clearance must still create custom, *ad hoc* methods to demonstrate equivalence or improved performance compared to existing systems. The lack of standardized evaluation methods introduces inefficiencies and inconsistencies in system characterization. While there has been growing support for using tissue-equivalent fluorescence phantoms to characterize system performance, variability in analysis methods limit reproducibility.[1–3] To address this gap, our work focuses on QUEL-QAL, an open-source software-driven analysis library that provides a foundational standardized approach for extracting quantitative performance metrics from fluorescence imaging systems.

Medical imaging has long relied on phantoms, physical models designed to evaluate system performance. Optical phantoms, a subset of these, simulate how light interacts with biological tissue, particularly its scattering and absorption properties.[4] These phantoms can take various forms, including liquids, gels, and solids. Recent advancements highlight the advantages of solid optical phantoms, which offer greater stability over time and benefit from emerging manufacturing techniques.

A key innovation in this space is the fluorescence reference target, a type of solid optical phantom embedded with fluorescent contrast agents.[5–7] These targets play a critical role in assessing fluorescence imaging systems, which are widely used in preclinical research, surgical guidance, and intraoperative pathological assessment. Our work introduces software solutions to automate and standardize quantitative analysis of fluorescence reference targets, ensuring that imaging system performance can be assessed consistently and objectively.

While the field is not yet standardized, there is growing consensus on the type of metrics needed to characterize imaging system performance, with the most definitive example being the efforts led by the American Association of Physicists in Medicine under Task Group 311.[1] This publication outlines the key quality metrics and collection methods needed for the evaluation of fluorescence-guided surgery systems. Others have also published on this topic,[8–12] and recent FDA draft publication on optical imaging guidance provides an appendix specific to device consideration.[13] A summary of the key metrics are provided in **Table 1**, where our current work focuses primarily on the characterization metrics in the first column.

While the AAPM TG311 report provides clear recommendations for the characterization metrics to report, the practical application of extracting this data from fluorescence images is not always trivial. Recent literature reports have demonstrated how commonly reported metrics can be significantly skewed by how target and background regions are selected.[2,3] The goal of this open-source image analysis library is both to provide an easier pathway for extracting quantitative metrics from fluorescence images, while increasing dialogue around the methods used, and ultimately improving reproducibility.

The following sections provide an overview of the QUEL-QAL library architecture. The three modules for region of interest extraction, statistical analysis, and visualization are described. The current work focuses on using the library for reference target analysis. While the library has functionality to perform analysis on anatomical visualization phantoms, discussion of this application will be outside the current scope. Example use cases are provided for reporting response linearity, limit of detection, depth sensitivity and image sharpness. The interested reader can explore more examples on the library wiki.

## Library Design

QUEL-QAL is an open-source Python library designed to analyze fluorescent targets in accordance with AAPM TG311 and FDA guidelines. The library supports the characterization of wide-field imaging systems operating in the near-infrared (NIR) fluorescent regime. It quantifies imaging performance metrics—such as fluorescence concentration sensitivity, fluorescence depth sensitivity, and spatial resolution—to help users understand and characterize the performance of their imaging systems.

Imaging in the NIR spectrum requires a different approach compared to visible light imaging. Since the human eye cannot distinguish wavelengths beyond the visible range, NIR fluorescence is best captured as a high-fidelity, monochrome image. This avoids the signal loss that can occurs when using Bayer-filtered RGB sensors optimized for visible light. Instead, dedicated optical filters are used to selectively sense NIR wavelengths, ensuring that the signal intensity is preserved for analysis.

In the sections that follow, we describe the methodologies implemented in QUEL-QAL for processing and analyzing various fluorescent targets, including concentration, depth, and resolution targets.

**Table 1: Key metrics used in fluorescence image system characterizations**

| System Characterization | | Application-Specific Validation |
|---|---|---|
| *Characterization Metrics* | *Imaging Device Settings* | *Clinical Indication* |
| Response linearity and dynamic range | Focal Length | Clinically meaningful limits of detection |
| Limit of Detection | Working Distance | Tissue absorption & scattering effects |
| Target to background ratio | Illumination wavelength | Crosstalk |
| Depth sensitivity | Detection Wavelength | Off-target fluorescence |
| Image sharpness (Resolution) | Illumination Intensity | Repeatability & reproducibility |
| Depth of Field | Binning | |
| Field of view | Gain | |
| Illumination Uniformity | Exposure time | |
| Detection Uniformity | Ambient light | |
| Distortion | | |
| Spatial co-registration | | |

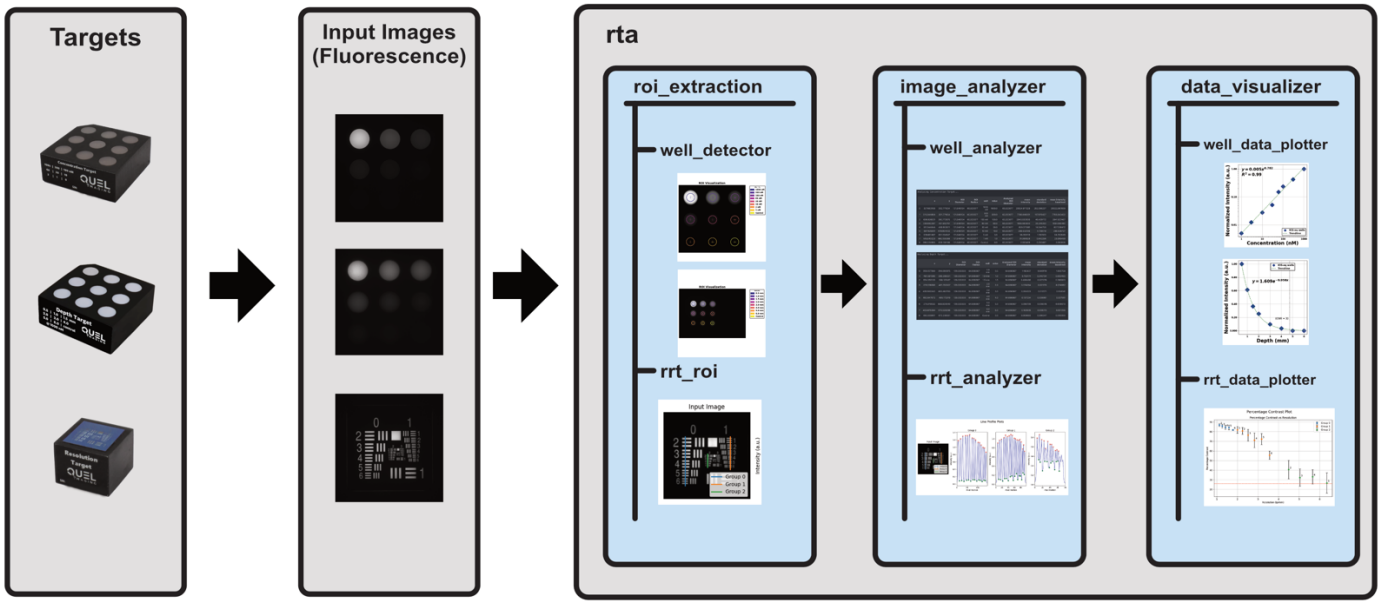

**Figure 1:** Graphical description of QUEL-QAL workflow. Reference Targets are used to collect fluorescence images. The QUEL-QAL code is use to identify regions of interest, extract statistics and visualize metrics.

QUEL-QAL is designed around a modular workflow that follows typical stages of fluorescence image analysis. The process is divided into three primary steps:

1. **ROI Extraction:** Identifying and defining region of interests (ROIs) in an image.
2. **Image Analysis:** Extracting quantitative data from each ROI.
3. **Data Visualization:** Presenting the analysis results in a clear, interpretable format.

The **ROI Extraction** module leverages libraries such as *NumPy*, *scikit-image*, *OpenCV*, *SciPy*, and *scikit-learn* to automate the detection and segmentation of fluorescent targets.[14–17] The goal is to accurately delineate the regions of interest while minimizing manual intervention. The **Image Analyzer** module performs statistical analysis of the intensity data contained in the identified ROIs. It computes key metrics—such as mean intensity, standard deviation, and contrast-to-noise ratios—to evaluate parameters like fluorescence concentration, depth sensitivity, and resolution. The **Data Visualizer** uses *Pandas* for data handling and *matplotlib* for plotting.[18–20] This module turns numerical outputs into informative graphs and charts. These visualizations help users quickly assess the performance of their fluorescence imaging systems.

By maintaining this clear separation of processes, QUEL-QAL not only simplifies the process of fluorescence analysis but also ensures that future extensions or adaptations of the library can follow the same established framework.

# Getting Started

QUEL-QAL is available on GitHub (https://github.com/QUEL-Imaging/quel-qal) and PyPI (*quel-qal*). It requires Python version 3.12 or newer. Examples and datasets are provided in the source code. The input images are expected to be monochrome fluorescent images, not white light color images or overlay images. Ideally, these images have had minimal processing applied to threshold or compress the original data.

While specific data collection techniques are not standardized, it is assumed the reference images will be collected normal to the reference target surface. Beyond this assumption, it is recommended to record the working distance and any pertinent image device settings, as described in **Table 1**.

To get started with QUEL-QAL, it is recommended to first create and activate a virtual environment before installing.

On macOS/Linux:

```
python3 -m venv env
source env/bin/activate
```

On Windows:

```
python -m venv env
env\Scripts\activate
```

QUEL-QAL can then be installed from PyPI using:

```
pip install quel-qal
```

The following is a simple Python script using QUEL-QAL to analyze an image of a 9-well fluorescence concentration target (the image is part of the QUEL-QAL examples dataset):

```
from qal.data import cn_sample_1
from qal import WellDetector, WellAnalyzer, WellPlotter

rcs_im = cn_sample_1()

detector = WellDetector()
detector.detect_wells(rcs_im)
rcs_df = detector.estimate_remaining_wells_3x3(
    rcs_im,
    detector.df,
    well_ids=['1000 nM','300 nM','100 nM','60 nM','30 nM','10 nM','3 nM','1
    nM','Control']
)

analyzer = WellAnalyzer(rcs_im, rcs_df)
rcs_df = analyzer.get_stats()

plotter = WellPlotter(rcs_df)
plotter.plot(
    graph_type='concentration',
    col_to_plot='mean intensity normalized',
    fluorophore_label='ICG',
    trendline_lib='statsmodels'
)
```

## Use Cases

Examples are provided using QUEL Imaging reference targets, but the same concept can be applied to other phantom and target designs. The following sections describe how the analysis is performed for reference targets containing a 3x3 grid of circular regions of interest. Circular ROIs are common in solid fluorescence target design due to manufacturability and potential for simulating simple geometric structures. The response linearity and limit of detection are extracted from a fluorescence concentration target. Depth sensitivity is measured from a target with similar shape, where the fluorescence concentration remains constant, but the thickness of overlying non-fluorescent material changes for each well. The image sharpness or resolution analysis is performed on a fluorescent target containing a USAF1951 test target.

# Response linearity, limit of detection and depth sensitivity

The concentration target is designed to characterize the fluorescence sensitivity of an imaging system. This target consists of nine circular wells arranged in a 3x3 matrix, where each well is of equal diameter and equally spaced. Each well contains a known concentration of a fluorophore (e.g., ICG) that decreases from the top-left to the bottom-right, with the bottom-right well serving as the control with tissue-mimicking properties and containing no fluorophore. In our examples, the fluorophore concentrations used are 0, 1, 3, 10, 30, 60, 100, 300, and 1000 nM, providing a broad dynamic range for assessing imaging sensitivity.[6]

The depth target evaluates the fluorescent signal of a fixed concentration embedded under varying thicknesses of tissue-mimicking material. Tissue-mimicking layers in the depth target include thicknesses of 0.5 mm, 1 mm, 1.5 mm, 2 mm, 3 mm, 4 mm, 5 mm, 6 mm, and a control well.[6]

While QUEL-QAL is agnostic to the specific fluorophore or concentration range, these examples assume a 3x3 structure and consistent well dimensions.

### ROI Extraction

The first step in analyzing the 3x3 target is to reliably detect the circular wells that constitute the target. Given the large dynamic range of the concentration target (1 to 1000 nM) and similar range of the depth target (imaging through 0.5 to 6 mm of tissue-equivalent material), traditional circle-detection methods are challenging. Instead of relying on methods such as blob detection or Hough transforms directly, QUEL-QAL employs a series of pre-processing and detection steps to optimize well identification.

**Pre-Processing and Detection Steps:** The image preparation assumes the input image is captured normal to the target top surface with minimal distortion, contains only one target, and is free of extraneous bright sources. Care must be taken during acquisition to avoid saturation in high-concentration wells and ensure that low-concentration wells remain above the noise floor. The bit-depth and contrast of the input image is modified to enhance the automation of well detection. This process involves normalizing the image and expanding the intensity values across the full 16-bit range, and converting the intensities to a logarithmic scale to enhance contrast between wells and background. A copy of the log-compressed image is then converted to a an 8-bit image. This conversion to an 8-bit range provides the ability to leverage existing image processing libraries (such as *OpenCV*), thereby reducing code complexity while reducing processing time. While these steps manipulate the image intensities, these changes are applied to a copy of the original image and only used for ROI identification.

**Rolling Window Thresholding:** Since all images are converted to 8-bit and their contrast is stretched over the entire dynamic range, a narrow window of intensity values (e.g., 245–255) is initially used to threshold the image, isolating regions that likely correspond to the brightest wells. The *scikit-image* function *measure.regionprops* is then applied to identify regions with low eccentricity (i.e., near-perfect circles), and key properties (position and area) of these regions are recorded. The intensity window is incrementally shifted (e.g., to 240–250, then 235-245, and so on) to capture wells at different intensity levels, ensuring that all wells are detected despite their varying brightness.

**Clustering and Refinement:** Due to overlapping intensity ranges across thresholds, the same well will be detected multiple times. To consolidate redundant detections, we apply agglomerative clustering using the implementation provided by scikit-learn,[17] grouping wells with similar centroid coordinates based on a distance threshold defined by the detected well diameter. Within each cluster, the average position is computed to define the final centroid, and all cluster diameters are normalized to the smallest detected diameter. This normalization mitigates the effects of blooming or noise artifacts. The resulting clusters are sorted by intensity and then labeled according to the predefined list of concentration values.

**Inference of undetected wells:** In cases where fewer than nine wells are detected, the known 3x3 grid layout of the concentration or depth target is used to infer the positions of the missing wells. This step is particularly important for reliably identifying wells with lower signal such as the control well, which might otherwise be lost in the noise floor. A minimum of three wells must be detected in the previous step for the inference to be run successfully. The *WellDetector.detect_wells* function is agnostic to the 3x3 well structure so other layouts would need to modify this inference function.

```
from qal.data import depth_sample_1
from qal import WellDetector, WellAnalyzer, WellPlotter
depth_im = depth_sample_1()

detector = WellDetector()
detector.detect_wells(depth_im)
depth_df = detector.estimate_remaining_wells_3x3(
    depth_im,
    detector.df,
    well_ids=['0.5 mm','1.0 mm','1.5 mm','2.0 mm','3.0 mm','4.0 mm','5.0 mm','6.0
    mm','Control']
)
```

**Output:** The final output of the ROI extraction process, implemented in the *WellDetector* class, is a *Pandas* dataframe. Each row in the dataframe contains:

- The x and y coordinates of the well's centroid,
- The ROI diameter and radius,
- A well identifier (ID) and its corresponding numerical value.

This structured output serves as the input for the *WellAnalyzer* class, which further computes quantitative metrics from each well's ROI.

## Well Analysis

Once the wells are identified and their regions of interest (ROIs) are extracted, the *WellAnalyzer* class quantifies key metrics from each ROI. These metrics are essential for evaluating the performance of the imaging system in terms of sensitivity and consistency.

By default, the analysis focuses on a region covering half the diameter of the full ROI, based on recommendations by AAPM TG311.[1] Analyzing this fractional region minimizes the effects of intensity drop-off and avoiding potential artifacts, thereby providing more robust and representative metrics. The size of the analyzed region can be adjusted using the *region_of_well_to_analyze* parameter in the *WellAnalyzer.get_stats* function, which specifies the fraction of the well to be analyzed.

```
analyzer = WellAnalyzer(depth_im, depth_df)
depth_df = analyzer.get_stats(region_of_well_to_analyze=0.5)
```

Specifically, the analysis calculates:

**Mean Intensity:** The average intensity within each well's ROI. This provides a baseline measure of fluorescence and helps detect issues such as overexposure or insufficient signal.

**Standard Deviation:** The variability of pixel intensities within the ROI. A low standard deviation typically indicates uniform illumination and consistent fluorophore distribution, while a high standard deviation may signal irregularities in the excitation source or non-uniform distribution of fluorophores in a well.

**Baselined Intensity:** The mean intensity for each well, adjusted by subtracting the intensity of the control (0 nM) well. This baseline correction accounts for any linearity issues stemming from a high noise floor, reflecting an accurate fluorescence response from an imaging system.

**Normalized Mean Intensity:** A relative measure computed by normalizing the well intensities ranging from the control well to the highest intensity well. This normalization enables the direct comparison of fluorescence responses across different concentrations, clarifying the linearity of an imaging system's response.

**Contrast-to-Noise Ratio (CNR):** The CNR is calculated using the following equation:

$$CNR = \frac{(\mu_{fl} - \mu_{bg})}{\sigma_{bg}}$$

Where $\mu_{fl}$ is the mean fluorescence intensity for each ROI, $\mu_{bg}$ is the mean fluorescence intensity of the background ROI (control well), and $\sigma_{bg}$ is the standard deviation of the fluorescence intensity in the background ROI. A CNR above 3 is considered the threshold for reliable detection,[1,21] indicating that the signal stands out sufficiently from the background of the image.

**Output:** The *WellAnalyzer* outputs a Pandas dataframe containing these calculated metrics for each well. This structured output not only facilitates further analysis but also serves as the input for subsequent visualization functions.

### Data Visualizer

The visualization module for the concentration target in QUEL-QAL is encapsulated in the *WellPlotter* class, which generates graphs using data from the previously generated dataframe. By leveraging Pandas for data handling and matplotlib for plotting, the *WellPlotter* class provides users with visual feedback on key performance metrics.

```
plotter = WellPlotter(depth_df)
plotter.plot(
    graph_type='depth',
    col_to_plot='mean intensity normalized',
    fluorophore_label='ICG-eq',
    trendline_lib='scipy',
    cnr_threshold=3
)
```

The primary input parameters of the data visualizer are described below:

**Graph Types:** Users can specify the type of graph to generate by passing the appropriate argument to the plot function (e.g., 'concentration' for concentration target visualizations or 'depth' for depth target analysis). This flexibility allows for consistent visualization across different target types.

**Log-Scaled Plots:** For concentration targets, the plots are rendered on a logarithmic scale along both axes. This approach enhances the visualization of the linearity of the imaging system's response over a wide dynamic range, making subtle variations in fluorescence intensity more apparent, especially on the lower end of fluorophore concentrations.

**Data Representation:** The primary datapoints plotted are the normalized mean intensities against the corresponding fluorophore concentrations. As the control well (0 nM) inherently has a normalized intensity of zero, it is excluded from the graph.

**Trendline and Statistical Metrics:** A trendline is automatically fitted to the data points that exhibit a contrast-to-noise ratio (CNR) greater than a user-defined threshold value (default of 3). The trendline, along with its coefficient of determination ($R^2$), is displayed on the graph. This statistical information is crucial for assessing the linearity and overall performance of the imaging system.

By integrating these visualization capabilities, QUEL-QAL not only streamlines the analysis workflow but also enhances the interpretability of the results, enabling users to make informed decisions about imaging system performance.

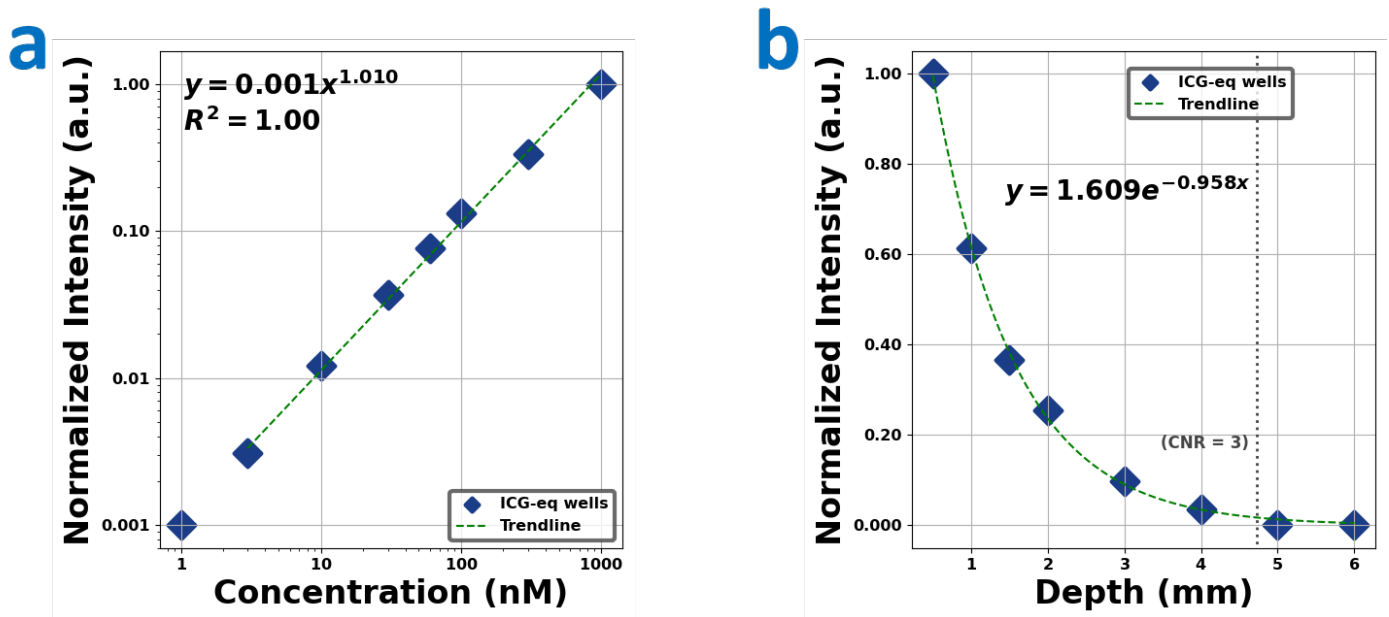

**Figure 2:** Example plots demonstrating (a) linearity characterization of a concentration sensitivity target with CNR > 3 at 3 nM ICG concentrations and greater, and (b) a depth sensitivity characterization showing a fluorescence depth sensitivity of ~4.7 mm.

# Image Sharpness

The QUEL resolution target uses a negative USAF 1951 resolution test chart with fluorescent backing, featuring groups 0 through 7. This is a standardized target for understanding spatial resolution of an optical imaging system. Each group contains 6 elements which have progressively smaller feature sizes. The elements consist of three lines. Determining the smallest resolvable group and element pair provides an estimate of the minimum feature size that the imaging system can resolve. Since the chart is standardized, the group and element can be correlated to spatial resolution in line pairs per mm (lp/mm) or line size (um). In the current library, Groups 0-2 are analyzed, which correlates to a spatial resolution of 1.12 - 7.13 lp/mm or 445 - 70 μm.

The analysis workflow for the resolution target mirrors the general pipeline used for other targets: ROI extraction, image analysis, and data visualization.

### ROI Extraction

The ROI extraction process for the resolution target takes a semi-autonomous approach:

**Automatic Cropping and Rotation:** The input image is automatically cropped and rotated to align the resolution target for further processing. This process uses template matching, keypoint detection, and homography transformation to detect and extract a resolution target (USAF 1951) from the input image, optionally saving the cropped image.

**Refining the ROI:** An interactive window prompts the user to select two key points:

- The top-left corner of the three horizontal bars in Group 0, Element 2.
- The bottom-right corner of the three horizontal bars in Group 0, Element 1.

Using these key points, scikit-image's corner_harris function is applied to accurately refine the key corner point for analysis.

**Line Profile Generation:** Once the key points are defined, vertical line profiles are automatically drawn along the fluorescent horizontal line pairs. These line profiles are generated for:

- Group 0: Elements 2–6,
- Group 1: Elements 1–6, and
- Group 2: Elements 2–6.

The start and end coordinates for each vertical line are calculated based on the known layout of the target and stored in a variable, *group_coordinates*.

```
from qal.data import res_sample_1
from qal import RrtROI, RrtAnalyzer, RrtDataPlotter

image = res_sample_1()

rrt_roi = RrtROI()
cropped_image = rrt_roi.get_resolution_target_cropped(image)
rrt_roi.select_points(cropped_image)
group_coordinates = rrt_roi.group_coordinates
```

**Resolution Target Analyzer**

Using the extracted *group_coordinates* as input, the analyzer computes several key metrics for each group and element, including:

**Resolution:** Reported in line pairs per millimeter.

**Line Width:** The width of the fluorescent lines in microns.

**Peak and Trough Positions:** The locations of local maxima (peaks) and local minima (troughs) along the generated line profiles.

**Percentage Contrast:** Calculated for each line pair using the formula:

$$\frac{(I_{max} - I_{min})}{(I_{max} + I_{min})} * 100$$

Where $I_{max}$ is the maximum fluorescence intensity or peak height of each element, and $I_{min}$ is the minimum intensity of the background. This metric is used in determining the resolvable features of each group-element.

**Standard Deviation of Percentage Contrast:** Used to assess the variability in contrast measurements.

```
analyzer = RrtAnalyzer()
contrast_df = analyzer.load_and_process_groups(cropped_image, group_coordinates)
```

The minimum resolvable feature is defined as the smallest group-element combination where the percentage contrast exceeds 26.4%, in line with the Rayleigh criterion.[1,8] The analyzer reports the contrast both in terms of line pairs per millimeter and the corresponding group-element.

**Data Visualizer**

The *ResolutionTargetPlotter* visualizes the analysis results by combining the original image, the extracted *group_coordinates*, and the dataframe containing the computed statistics:

```
visualizer = RrtDataPlotter()
visualizer.plot_line_profiles(cropped_image, group_coordinates, percentage_contrast_df)
visualizer.plot_percentage_contrast(percentage_contrast_df)
```

In this example, elements in groups 0-2 are analyzed by identifying the peaks ($I_{max}$) and troughs ($I_{min}$) along each groups' line profile. Figure 3 provides a visual example of this process where the left panel displays the resolution target image with the overlay of the vertical line profiles used in the analysis for group 0, group 1, and group 2. The line graphs on the right show a plot for each of the three groups that line profiles were generated for. Each plot contains the intensity values plotted against the position of the line profiles. Markings of the pixel positions denote the local maxima (peaks) and local minima (troughs).

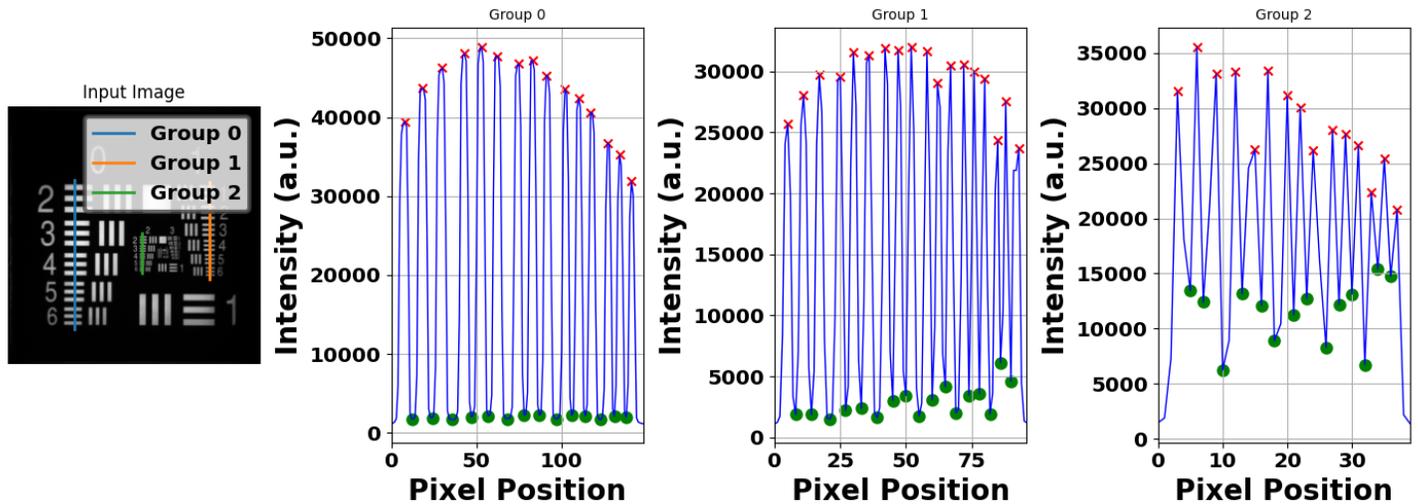

**Figure 3: Visual workflow of how the resolution target is analyzed to extract percent contrast across groups 0, 1 and 2.**

In each element, the arrangement comprises three fluorescent lines alternating with two non-fluorescent lines. As the feature size is reduced, the peak intensity falls, while at the same time the background intensity increases. Fundamentally this is due to signal blurring at the pixel level and distinguishing between fluorescence signal and background becomes more challenging. The Rayleigh criterion states when the contrast drops below 26.4% the resolution limit has been reached. In the example provided in **Figure 4**, this threshold is crossed in Group 2 between elements 5 and 6, or ~7 lp/mm.

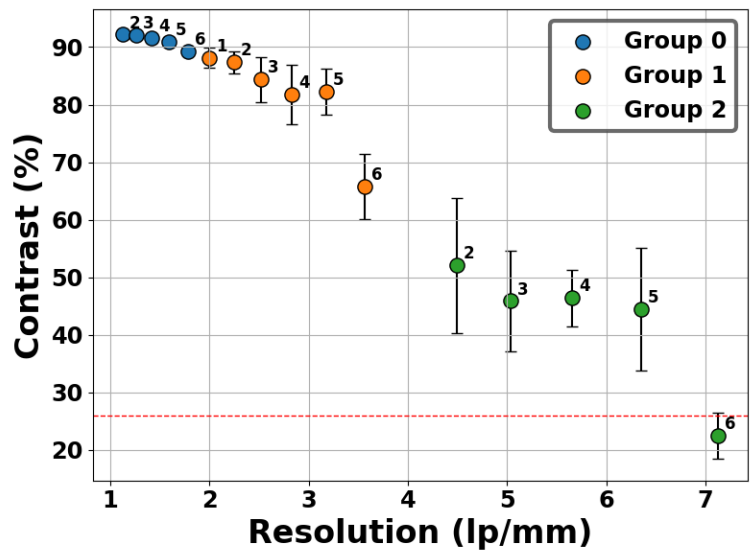

**Figure 4: Example of how the percent contrast can be extracted from an image of the resolution target to determine optimal spatial resolution.**

# Discussion

Previous groups have reported using template matching with solid fluorescence phantoms for benchmarking system performance, however, this code required a MATLAB subscription to run.[22,23] Additionally, earlier iterations of the current work were available through a web-portal interface, but this mechanism required users to upload imaging data and the analysis methods were not available for external review. Industry users were reluctant to upload images from systems under development, and the black-box analysis did not aid in FDA regulatory reviews.

Other benchmarking methods have highlighted the importance of background selection in the analysis process.[2,3,21] By manipulating how the background is defined, the overall quantitative analysis can be skewed, which make meta-analysis between studies nearly impossible. By implementing an extensible Python framework, available on GitHub and PyPI, we aim to allow users the flexibility to perform their own analysis, using version-controlled analysis methods, which should accelerate their FDA regulatory reviews.

With growing adoption, we plan to use QUEL-QAL, and the included example data, as a framework to discuss best practices in image analysis and system characterization. We hope these discussions can lead to contributions to develop common benchmarking methods and improve reproducibility across the field. The provided use cases demonstrate the foundation of the QUEL-QAL functionality. Additional examples describing methods to measure uniformity and distortion are available on GitHub.

Beyond system characterization using reference targets, QUEL-QAL provides functionality to analyze fluorescence anatomical phantoms. This functionality is currently being expanded to support phantoms demonstrating indication-specific applications, such as ex vivo margin assessment. Our goal is to expand the utility of tissue-equivalent fluorescence phantoms by standardizing how imaging metrics are collected and reported. The versatility of the Python library augments the value of phantoms by providing an efficient method to extract data and improve reproducibility and ultimately accelerate the pathway to standardization.

# Conclusions & Future Work

The current work provides a foundation to build consensus and work toward standardization in the field of fluorescence guided surgery. While the AAPM TG311 report and FDA draft guidance provide recommendations for system characterization metrics, the practical data collection and extraction can be cumbersome and prone to error or user interpretation. QUEL-QAL is an open-source library designed to expedite data extraction from common solid phantom designs. By utilizing a version-controlled library, end users can easily develop standardized workflows built on concurrent best practices. While the examples provided in this work focus on a set of commercially available reference targets, the extensible software framework provides the ability to easily add new features. There is on-going work to support new reference target designs as well as anatomical visualization phantoms. Future work will focus on new applications for the code and utilizing this framework to build consensus on best practices and move the field to adopt standardized practices.

# Acknowledgements

The QUEL-QAL library was funded in part with federal funds from National Cancer Institute, National Institutes of Health, and Advanced Research Projects Agency for Health, Department of Health and Human Services, under Contract Numbers 75N91021C00035 and 75N91023C00052. The reference targets used in this study were developed under SBIR grant from the National Institute of Biomedical Imaging and Bioengineering, National Institutes of Health, Department of Health and Human Services, under grant number R43/44EB029804.

# Disclosures

E LaRochelle is co-founder and owner of QUEL Imaging. All authors are full-time employees of QUEL Imaging.